\documentclass[12pt]{article}
\usepackage{graphicx}
\usepackage{amsmath,amsthm,amsfonts,amssymb,bm}
\usepackage{color,url}
\makeatletter
\@addtoreset{equation}{section}

\makeatother

\begin{document}
\begin{titlepage}
\null
\begin{flushright}
\end{flushright}

\vskip 1.5cm
\begin{center}

 {\LARGE \bf New Soliton Solutions of}

\vskip 0.6cm

 {\LARGE \bf Anti-Self-Dual Yang-Mills Equations}

\vskip 1.5cm
\normalsize

{\large 
Masashi Hamanaka\footnote{E-mail:hamanaka@math.nagoya-u.ac.jp}
and 
Shan-Chi Huang\footnote{E-mail:x18003x@math.nagoya-u.ac.jp}
}

\vskip 0.5cm

        { Department of Mathematics, University of
 Nagoya,\\
                      Nagoya, 464-8602, JAPAN}

\vskip 1.5cm

        {\large \it Dedicated to the memory of Jon Nimmo}

\vskip 1.5cm

{\bf \large Abstract}

\vskip 0.5cm
 
\end{center}

We study exact soliton solutions of anti-self-dual Yang-Mills 
equations for $G =GL(2)$ in four-dimensional spaces with 
the Euclidean, Minkowski and Ultrahyperbolic signatures
and construct special kinds of one-soliton solutions whose
action density Tr$F_{\mu\nu}F^{\mu\nu}$ can be real-valued.
These solitons are shown to be new type of domain walls in four dimension 
by explicit calculation of the real-valued action density. 
Our results are successful applications of the Darboux transformation 
developed by Nimmo, Gilson and Ohta. 
More surprisingly, integration of these action densities 
over the four-dimensional spaces 
are suggested to be not infinity but zero. 
Furthermore, whether 
gauge group $G= U(2)$ can be realized on our solition solutions 
or not is also discussed on each real space.


\end{titlepage}

\clearpage
\baselineskip 6.22mm


\section{Introduction}

Yang-Mills theories are at the center of
elementary particle physics to describe
fundamental laws of interactions.
Topological solitons in these theories, 
such as instantons, monopoles, vortices, calorons, merons,
played central roles in the study of
non-perturbative aspects, duality structures, quark confinements and so on.
(See e.g. \cite{Actor,Coleman,Monopole,DHKM,GPY,MaSu,Polyakov,Shifman,tHooft}.)
To study these topological solitons, 
the anti-self-dual (ASD) Yang-Mills equation
would be in the most important position. For instance, 
the instantons are global solutions of this equation
with a special boundary condition such that the action is finite.
For mathematical aspects, the instantons are
described very elegantly by the ADHM construction \cite{ADHM}.  

On the other hand, the anti-self-dual Yang-Mills equation 
has a very close relationship
with lower-dimensional integrable equations, such as the KdV equation,
the Toda equations, the Painlev\'e equations and so on \cite{MaWo, Ward}.
Energy densities of some soliton solutions to these equations 
are localized on hyperplanes in the whole space-time dimensions 
and hence they can be interpreted as domain walls in the space-times.
Existence of these solitons solutions also relate to
their integrability, such as  existence of infinite
conserved quantities and existence of hidden infinite symmetries.
For anti-self-dual Yang-Mills equations,
the domain wall type soliton solutions exist as well and can be constructed
from the 't Hooft ansatz and the Atiyah-Ward ansatz. 
However, already known soliton solutions 
given in section 4 always lead to trivial action densities as we will see.

In this paper, we construct exact soliton solutions
of anti-self-dual Yang-Mills equations for $G = GL(2)$  
and calculate the action densities of them on four-dimensional real spaces with
the Euclidean, Minkowski and Ultrahyperbolic signatures.
We find that these type Soliton solutions lead to 
real-valued action densities which can be interpreted as 
non-trivial domain walls in four-dimension.
This beautiful result is a successful application 
of the Darboux transformation
developed by Nimmo, Gilson and Ohta \cite{GNO}.
More surprisingly, integration of these non-trivial action densities
over the four-dimensional spaces are 
suggested to be
not infinity but zero.
We also discuss in details whether 
gauge group could be unitary on our solition solutions or not and find that $G=SU(2)$ could be realized 
in one kind of the Ultrahyperbolic signature. 

This paper is organized as follows.
In section 2, we introduce the anti-self-dual Yang-Mills equations
on four-dimensional complex spaces and give exact soliton solutions
together with action densities of them.  
In section 3, we present exact soliton solutions with real valued action densities 
by taking some dimensional reduction conditions on the complex spaces 
and discuss the possibility of realization of unitary gauge group on each real space.
In section 4, we review some already known soliton solutions of
the anti-self-dual Yang-Mills equations and show that they are all trivial
in the sense of action density while our solutions are non-trivial.
Section 5 is devoted to conclusion and discussion.

\section{Soliton solutions on four-dimensional complex spaces}

In this section, we give a complex version of 
four-dimensional anti-self-dual Yang-Mills equations 
which is a unified treatment of section 3. 
In section 2.1, 
we introduce a formulation of anti-self-dual Yang-Mills equations 
on four-dimensional complex spaces which relates to the twistor theory, following the conventions close to that in the book 
of Mason and Woodhouse \cite{MaWo}.
In section 2.2, 
we calculate a complex-valued action density of exact soliton solutions \cite{GHHN} 
generated by the Darboux transformation \cite{GNO}.
This complex-valued action density would be reduced to
four-dimensional real spaces with three kinds of signatures in section 3 
and the reduced action densities could be real-valued by taking some conditions.

\subsection{Anti-Self-Dual Yang-Mills Equations}

Let $(z,\widetilde z, w, \widetilde w)$ be 
a double null coordinates on four-dimensional complex spaces
with metric defined by
\begin{eqnarray}
ds^2&=&g_{mn}dz^{m}dz^{n}
	=2(dz d\widetilde z -dw d\widetilde w),
~~~m,n=1,2,3,4.\\
&&{\mbox{where}}~~
\label{N=2_com}
g_{mn}:=\left(
\begin{array}{cccc}
0&1&0&0 \\
1&0&0&0 \\ 
0&0&0&-1 \\
0&0&-1&0 
\end{array}\right),~
(z^{1},z^{2},z^{3},z^{4}):=(z,\widetilde z, w, \widetilde w).\nonumber
\end{eqnarray}

We can recover three kinds of real spaces  
by taking some suitable reality conditions 
on $z,\widetilde z, w, \widetilde w$ as follows.
Concrete realizations are given in section 3. 
\begin{itemize}

 \item Reality condition:$\widetilde z=\overline z,~
\widetilde w= -\overline w$ gives 
the Euclidean real space $\mathbb{E}$.

 \item Reality condition:$z, 
\widetilde z \in \mathbb{R},~\widetilde w= \overline w$
gives the Minkowski real space $\mathbb{M}$.

 \item Reality condition:$\widetilde z=\overline z,~\widetilde w= \overline w$
gives the Ultrahyperbolic real space $\mathbb{U}_1$.  

\item Reality condition:$z, \widetilde z,w, \widetilde w\in \mathbb{R}$
gives the Ultrahyperbolic real space $\mathbb{U}_2$.
\end{itemize}
Note that $\mathbb{U}_1$ and $\mathbb{U}_2$ are different real slices 
even though their signature are the same.\\

Let us consider a gauge theory on the complex space and assume 
gauge group to be $G=GL(N)$. 
The field strengths are defined by 
\begin{eqnarray}
 F_{mn}:= \partial_{m} A_{n}- \partial_{n} A_{m}+[A_m, A_n], 
\end{eqnarray}
where $A_m(z)$ denote gauge fields 
which take values in the Lie algebra of $G$. 
The anti-self-dual Yang-Mills equation on the complex space is defined 
as follows: 
\begin{eqnarray}
\label{asdym_cpx}
F_{zw}=0,~ F_{\widetilde z \widetilde w} =0,~ 
F_{z \widetilde z}-F_{w \widetilde w} =0,
\end{eqnarray}
which reduces to the standard anti-self-dual Yang-Mills equations 
on real slices in the sense of Hodge dual as we will see in section 3.

In order to find the solution of the anti-self-dual Yang-Mills equations,  
let us begin with the Yang equation:
\begin{eqnarray}
\label{Yang}
\partial_{\widetilde z}(J^{-1}\partial_{z}J)-
\partial_{\widetilde w}(J^{-1}\partial_{w}J)=0,
\end{eqnarray}
where the $N\times N$ matrix is called Yang's $J$-matrix.
Then ASD gauge fields could be obtained from a solution $J$ 
of the Yang equation by decomposing $J$ into 
$N\times N$ two matrices $h$ and $\widetilde{h}$ 
such that $J = \widetilde{h}^{-1}h$~\footnote{
Note that the relation between $J$ and $h$, $\widetilde{h}$
is different from $J:=\widetilde{h}h^{-1}$ in \cite{GHHN}.}
 ~and setting:
\begin{eqnarray}
\label{ASD_gauge_f}
A_{z}=-(\partial_{z} h) h^{-1},~ A_{w}=-(\partial_{w} h) h^{-1},~
A_{\widetilde z}= -(\partial_{\widetilde z} \widetilde h) \widetilde h^{-1},~
A_{\widetilde w}= -(\partial_{\widetilde w}\widetilde h) \widetilde h^{-1}.
\end{eqnarray}
Note that the gauge transformation acts on 
$h$ and $\widetilde{h}$ as $h\mapsto gh,~\widetilde h\mapsto g\widetilde h,~
g(x)\in G$ and hence Yang's matrix $J$ is gauge invariant. 
If we take a special gauge $\widetilde{h}=1$, 
gauge fields become a simpler form 
in terms of $J$:
\begin{eqnarray}
\label{gauge_f_special}
A_{z}=J^{-1}\partial_{z}J,~~A_{w}=J^{-1}\partial_{w}J,~~
A_{\widetilde z} = A_{\widetilde w} = 0,
\end{eqnarray}
and satisfy the anti-self-dual Yang-Mills equation. 
Hence, we can define 
the following quantity and called it  
the action density in this paper:
\begin{eqnarray}
{\mbox{Tr}}F^2:={\mbox{Tr}}F_{mn}F^{mn}
=
-2{\mbox{Tr}}(
{{F}_{w{\widetilde{w}}}^2}+{{F}_{z{\widetilde{z}}}^2}
+2F_{{\widetilde{z}}w} F_{z{\widetilde{w}}}
+2F_{zw} F_{{\widetilde{z}}{\widetilde{w}}}
),
\end{eqnarray} 
where $F^{mn}:=g^{mk}g^{nl}F_{kl}$.
For ASD gauge fields, ${\mbox{Tr}}F^2=
4{\mbox{Tr}}(F_{w{\widetilde{z}}} F_{z{\widetilde{w}}}
-{{F}_{w{\widetilde{w}}}^2})$.

\subsection{Soliton Solutions and Action Densities for $G=GL(2)$}

From now on, let us focus on  
soliton solutions for $G=GL(2)$
generated from a trivial seed solution 
$J=1$ by the Darboux transformation \cite{GNO}.


The following $2\times 2$ complex matrix $J$
is a solution of the Yang equation \cite{GNO}.
	\begin{eqnarray}
	\label{general_sol}
	J=-Q\Lambda^{-1}Q^{-1},~~~
	\end{eqnarray}
	where $\Lambda$ is a constant $2\times 2$ matrix
	and $Q$ is a $2\times 2$ matrix satisfying
	\begin{eqnarray}
	\label{chasing}
	\partial_w Q=(\partial_{\widetilde{z}}Q) \Lambda,~
	\partial_z Q=(\partial_{\widetilde{w}}Q) \Lambda.
	\end{eqnarray}
	Soliton solutions are given
	by setting $Q$ and $\Lambda$ as follows \cite{GHHN}:
\begin{eqnarray}
\label{soliton_cpx}
&&Q=\left(
\begin{array}{cc}
a_1e^{L}+a_2e^{-L}
&
b_1e^{M}+b_2e^{-M}
\\
c_1e^{L}+c_2e^{-L}
&
d_1e^{M}+d_2e^{-M}
\end{array}\right),~~~
\Lambda=
\left(
\begin{array}{cc}
\lambda & 0 \\
0 & \mu
\end{array}
\right),\\
\label{LM_cpx}
&&L:=\lambda \beta z
+\alpha\widetilde{z}
+\lambda \alpha w
+\beta\widetilde{w},~~
M:=\mu \delta z
+\gamma\widetilde{z}
+\mu\gamma w
+\delta\widetilde{w},
\label{dispersion_cpx}
\end{eqnarray}
where $a_{1}, a_{2},
b_{1}, b_{2},
c_{1}, c_{2},
d_{1}, d_{2},
\alpha, \beta, \gamma, \delta, \lambda, \mu$
are complex constants.
Note that we only consider this type of solution in this paper from now on, that is, $J$ and $Q$ in \eqref{general_sol} and \eqref{soliton_cpx}, respectively.
 
After a little bit lengthy calculation, 
we can obtain explicit form of the action density 
with respect to this solution (For the details, see Appendix.):
\begin{eqnarray}
\label{action_cpx}
{\mbox{Tr}} F^2=
8(\lambda-\mu)^2(\alpha\delta-\beta\gamma)^2
\varepsilon_0\widetilde{\varepsilon}_0 
	\left[\frac{2\varepsilon_1\widetilde{\varepsilon}_1
		\sinh^2 X_1
		-2\varepsilon_2\widetilde{\varepsilon}_2
		\sinh^2 X_2 
		-\varepsilon_0\widetilde{\varepsilon}_0}
	{\left((\varepsilon_1\widetilde{\varepsilon}_1)^{\frac12}
		\cosh X_1 
		+(\varepsilon_2\widetilde{\varepsilon}_2)^{\frac12}
		\cosh X_2\right)^4}
	\right]	
\end{eqnarray}
where
\begin{eqnarray}
&&X_1:=M+L+\dfrac12 \log(\varepsilon_1/\widetilde{\varepsilon}_1),~~~
X_2:=M-L+\dfrac12 \log(\varepsilon_2/\widetilde{\varepsilon}_2)\\
&&\varepsilon_0:=a_2c_1-a_1c_2,~~~
\widetilde{\varepsilon_0}:=b_2d_1-b_1d_2,~\\
&& 
\varepsilon_1:=a_1d_1-b_1c_1,~~~
\widetilde{\varepsilon}_1:=a_2d_2-b_2c_2,~\\
&&
\varepsilon_2:=a_2d_1-b_1c_2,~~~
\widetilde{\varepsilon}_2:=a_1d_2-b_2c_1.
\label{epsilon2}
\end{eqnarray}
        Note that the action density
		vanishes identically when $\lambda=\mu$ or $\alpha\delta=\beta\gamma$ or
		$\varepsilon_{0}\widetilde{\epsilon_{0}}=0$.
		Furthermore, 
		$\varepsilon_{1}\widetilde{\epsilon_{1}}=\varepsilon_{2}\widetilde{\epsilon_{2}} \iff \varepsilon_{0}\widetilde{\epsilon_{0}}=0$. 
	This means the singularities appear on the locus $D:=\left\{(z, \widetilde{z}, w,\widetilde{w})\in\mathbb{C}^4~|~\
	(\varepsilon_1\widetilde{\varepsilon}_1)^{\frac12}
	\cosh X_1 
	+(\varepsilon_2\widetilde{\varepsilon}_2)^{\frac12}
	\cosh X_2=0,~
	\varepsilon_1\widetilde{\varepsilon}_1 \neq \varepsilon_2\widetilde{\varepsilon}_2
	\right\}$ 
	and $D$ is clearly nonempty because 
$X_I=i\left(n_I+1/2\right)\pi~(I=1,2,~n_I\in\mathbb{Z})$
satisfies $\cosh X_1=\cosh X_2=0$. 
To find other singularities, for example, we can impose a simple constraint: 
$\varepsilon_1 = k \varepsilon_2,~
\widetilde{\varepsilon_1} = k \widetilde{\varepsilon_2},~ 
k \in \mathbb{R} \setminus \left\{-1, 0, 1 \right\} $ 
satisfying the condition $\varepsilon_1\widetilde{\varepsilon}_1 \neq \varepsilon_2\widetilde{\varepsilon}_2$ so that phase shift factor
$1/2\log(\varepsilon_1/\widetilde{\varepsilon}_1)
=1/2\log(\varepsilon_2/\widetilde{\varepsilon}_2)=:\phi$.
Then some classes of singularities would appear on the sub-locus of D~:
\begin{eqnarray}
\widetilde{D}_1:=\left\{(z, \widetilde{z}, w,\widetilde{w})\in\mathbb{C}^4~|~\tanh L\tanh(M+\phi)=-(|k|+1)/(|k|-1)
\right\},~
\end{eqnarray}
 and a special class of them can be found explicitly on the intersection of complex hyperplanes defined by
$L=i\left\{\arctan ((|k|+1)/(|k|-1))+n_{L}\pi\right\}$
~and~ 
$M=i(n_{M}+1/4)\pi-\phi$~$(n_L, n_M \in \mathbb{Z})$. 
Next let us consider another example of sub-locus of $D$ :
\begin{eqnarray}
\label{locus_2}
\widetilde{D}_{2}:= D \cap 
\left\{ {(z, \widetilde{z}, w,\widetilde{w})\in\mathbb{C}^4~|~\mbox{Im}}X_I = n_I \pi, n_I \in \mathbb{Z}, I=1, 2
\right\},
\end{eqnarray}
which would greatly simplify the problem of locus $D$ to a homogeneous system of two linear equations of $\cosh({\mbox{Re}}X_I)$ by the argument formula : $\cosh X_I=\cosh({\mbox{Re}}X_I+i n_I \pi)
=(-1)^{n_I}\cosh({\mbox{Re}}X_I)$.
Then under the condition:
$\varDelta:=\left|
\begin{array}{cc}
{\mbox{Re}}(\varepsilon_1\widetilde{\varepsilon}_1)^{1/2} & 
{\mbox{Re}}(\varepsilon_2\widetilde{\varepsilon}_2)^{1/2}  \\
{\mbox{Im}}(\varepsilon_1\widetilde{\varepsilon}_1)^{1/2} & 
{\mbox{Im}}(\varepsilon_2\widetilde{\varepsilon}_2)^{1/2}
\end{array}
\right| \neq 0$, the singularities of action density are removed successfully since the singular sub-locus $\widetilde{D}_{2}$ becomes empty.

Classifying all the singularities of the complex action density \eqref{action_cpx} in details is a rewarding job, however, we would like to discuss this issue in a separated paper because our aim in this paper is to study the real-valued action density for physical purpose. In fact, we can exclude all the singularities on each real slice by adjusting parameters $a_I,b_I,c_I,d_I$ in \eqref{soliton_cpx} 
and taking reality conditions on \eqref{LM_cpx}. The remaining details are discussed in chapter 3.

On the other hand, we also find that the action density has two principal peaks lie on ${\mbox{Re}}X_1=0$ and ${\mbox{Re}}X_2=0$ 
on the no singularity region, as we mentioned in \eqref{locus_2}.
This fact is very interesting and quite different 
from our experience in the lower-dimensional soliton equations. 
More precisely, the principal peaks of soliton configurations of lower-dimensional soliton equations usually lie on the ${\mbox{Re}}L=0$ or ${\mbox{Re}}M=0$ (up to phase shift factors), however, our principal peaks lie on ${\mbox{Re}}(M\pm L)=0$. 
Let us consider the same analysis of the
anti-self-dual Yang-Mills equation, 
like that for lower-dimensional solition equations. Firstly, we 
take a limit of 
$r^2:=\vert \sum_{m=1}^4 z^m\overline{z}^m\vert^2
\rightarrow \infty$ so that $\vert L \vert$ is finite
in the solution \eqref{soliton_cpx}.  
Then $\vert e^{M}\vert$ goes to infinity or zero
and $\vert e^{-M}\vert$ goes to
zero or infinity, respectively. That is,
\begin{eqnarray}
Q\stackrel{r^2\rightarrow\infty}
{{\longrightarrow}}
\left(
\begin{array}{cc}
a_1e^{L}+a_2e^{-L}
&
b_1e^{M}
\\
c_1e^{L}+c_2e^{-L}
&
d_1e^{M}
\end{array}\right)
{\mbox{~~or~~}}
\left(
\begin{array}{cc}
a_1e^{L}+a_2e^{-L}
&
b_2e^{-M}
\\
c_1e^{L}+c_2e^{-L}
&
d_2e^{-M}
\end{array}\right).
\end{eqnarray}
Note that the former and latter cases correspond to
$(b_2,d_2)=(0,0)$ and $(b_1,d_1)=(0,0)$, respectively.
By comparing \eqref{action_cpx},
the resulting action density vanishes in the both cases
while in the lower-dimensional soliton equations,
the configuration has its principal peak on ${\mbox{Re}}L=0$ by similar analysis.

Inspired from the above analysis, let us focus on 
one principal peak of \eqref{action_cpx} and set $a_2=b_1=c_1=d_2=0$. 
Then we can obtain a reduced form of our soliton solution
\begin{eqnarray}
\label{1soliton_cpx}
Q=\left(
\begin{array}{cc}
ae^{L}
& 
be^{-M}
\\ 
ce^{-L}
& 
de^{M} 
\end{array}\right),~~~a,b,c,d\in\mathbb{C},
\end{eqnarray}
which leads to a simpler form of action density: 
\begin{eqnarray}
\label{action_cpx_1}
{\mbox{Tr}} F^2=
8(\lambda-\mu)^2(\alpha\delta-\beta\gamma)^2
\left(
2{\mbox{sech}}^2 X-3{\mbox{sech}}^4 X
\right),	
\end{eqnarray}
where 
$X:=M+L+\dfrac12 \log(-ad/bc)$. 
(Note that $\varepsilon_0 \widetilde{\varepsilon}_0 =
\varepsilon_1 \widetilde{\varepsilon}_1 = -abcd,~
\varepsilon_2 \widetilde{\varepsilon}_2 = 0$.)

Now let us discuss the singularity problem of the reduced action density \eqref{action_cpx_1} by the following argument formula of hyperbolic functions :
\begin{eqnarray}
\label{sech2}
\mbox{sech}^2(x+iy)
=2\left[
\frac{\mbox{cosh}2x ~\mbox{cos}2y +1}
{(\mbox{cosh}2x + \mbox{cos}2y)^2}
-i\frac{\mbox{sinh}2x~\mbox{sin}2y}{(\mbox{cosh}2x + \mbox{cos}2y)^2}
\right].
\end{eqnarray}
We find that \eqref{action_cpx_1} has 
periodicity on the slice spaces : $X= a + i~{\mbox{Im}}X$ for any given real number $a$, and the singularities appear periodically in the case of $a=0$ because ${\mbox{sech}}^2X={\mbox{sec}}^2({\mbox{Im}}X)$ if $X = i~{\mbox{Im}}X$ . Therefore, \eqref{action_cpx_1} has no solitonic behavior on slice spaces if the real part of X is fixed. On the other hand, 
to remove the singularities and periodicity,   
we can impose some constraint on the imaginary part of $X$.  
For example, taking the condition $X={\mbox{Re}}X + in\pi$ for $n \in \mathbb{Z}$ 
would achieve this goal as the following: 
\begin{eqnarray}
\label{action_cpx_2}
{\mbox{Tr}} F^2=
8(\lambda-\mu)^2(\alpha\delta-\beta\gamma)^2
\left(
2{\mbox{sech}}^2 ({\mbox{Re}}X)-3{\mbox{sech}}^4 ({\mbox{Re}}X)
\right), 
\end{eqnarray}
which possesses real-valued solitonic behavior up to complex constants. 
For other nontrivial examples, 
we can use formula \eqref{sech2} and consider the slice spaces~: $X = {\mbox{Re}}X + i(n\pm 1/4)\pi$ for $n \in \mathbb{Z}$ to get the result~:
\begin{eqnarray}
\label{action_cpx_3}
\begin{split}
2{\mbox{sech}}^2 X-3{\mbox{sech}}^4 X
&=
8\left(
2{\mbox{sech}}^2 (2{\mbox{Re}}X)-3{\mbox{sech}}^4 (2{\mbox{Re}}X)
\right)\\
&~~~\pm 4i(6{\mbox{sech}}(2{\mbox{Re}}X)-1)
{\mbox{sech}}^2 (2{\mbox{Re}}X)
{\mbox{tanh}}(2{\mbox{Re}}X).
\end{split}
\end{eqnarray}
Note that \eqref{action_cpx_3} which belongs to a new class 
of solutions, is quite different from \eqref{action_cpx_2} because 
the solitonic behavior appears in both real part and imaginary part.

Finally, we remark a condition for $J$-matrix such that $J$ is unitary. 
We hope that our understanding of $J$-matrix would be helpful for the realization of $G=U(N)$ since the action density becomes real-valued and fit to physical interpretation when $G=U(N)$.
Let us put a condition on the solution 
$Q$ in \eqref{general_sol} as follows
\begin{eqnarray}
\label{ps_unitary}
Q=\left(
\begin{array}{cc}
A & B \\ 
-\overline B & \overline A
\end{array}\right),~~~
\Lambda=
\left(
\begin{array}{cc}
\lambda & 0 \\
0 & \mu
\end{array}
\right).
\end{eqnarray}
Then Yang's $J$ matrix becomes
\begin{eqnarray}
\label{}
J=\frac{-1}{\left| A \right|^2+\left| B \right|^2}
\left(
\begin{array}{cc}
(1/\lambda)\left| A \right|^2+(1/\mu)\left| B \right|^2
& (1/\mu-1/\lambda)AB \\ 
(1/\mu-1/\lambda)\overline A \overline B
& (1/\mu)\left| A \right|^2+(1/\lambda)\left| B \right|^2
\end{array}\right).
\end{eqnarray}
Hence, we can find that under the condition \eqref{ps_unitary},
$J\in U(2)\Leftrightarrow\left|\lambda\right|=\left|\mu\right|=1$ and 
$J\in SU(2)\Leftrightarrow\mu=\overline \lambda$ ~and~ $\left|\lambda\right|=1$.

We will see it soon in the next chapter that the ansatz \eqref{ps_unitary} gives a magical way to construct ${\mbox{Im}}X=0$ type action densities \eqref{1action_E}, \eqref{1action_M}, \eqref{1action_U1}, and \eqref{1action_U2} belonging to the same class ($n=0$) of \eqref{action_cpx_2}.

\section{Soliton Solutions on Four-dimensional Real Spaces}

In this section, we construct 
soliton solutions on four-dimensional real spaces
with three kinds of signatures and
the corresponding action densities 
could be realized to real-valued functions 
by taking the reality conditions in section 2.1
and condition \eqref{ps_unitary}. 
More precisely,
$c_1=-\overline{b}_1, c_2=-\overline{b}_2, 
d_1=\overline{a}_1,d_2=\overline{a}_2$
and $\overline{M}=L$. 
The latter condition gives rise to
relations between parameters $\alpha,\beta,\gamma,\delta,\lambda,\mu$  
on each real slice. 
After these replacements, the action density 
Tr$F^2$ reduces to the standard one: 
Tr$F_{\mu\nu}F^{\mu\nu}$
with respect to local coordinates $x^\mu~(\mu=0,1,2,3)$ on the four-dimensional real spaces. We can even show that the action density Tr$F_{\mu\nu}F^{\mu\nu}$ is real-valued because $X_1$ becomes 
real and $X_2$ becomes pure imaginary. 

More interestingly, the soliton solutions \eqref{1soliton_cpx} represent domain wall solutions and the
integration of the corresponding action densities over the real spaces 
are suggested to be infinity but zero. We put the proof in section 3.1. 
This property might shed light on a new study area of domain walls in cosmology. 

On the other hand, we will see that $G=U(N)$ can be realized only on the Ultrahyperbolic space $\mathbb{U}_2$ in section 3.4 because both gauge fields $A_\mu$ and field strengths $F_{\mu\nu}$ must take values in anti-hermitian $N\times N$ matrices when $G=U(N)$.

\subsection{On Euclidean Real Space $\mathbb{E}$}

To realize the Euclidean real slice condition: 
$\widetilde z=\overline z,~\widetilde w= -\overline w$,
we take the following combination of the real coordinates
$x^0,x^1,x^2,x^3$ on $\mathbb{E}$:
\begin{eqnarray}
\label{E} 
z=\frac{1}{\sqrt2}(x^0-ix^1),~ 
\widetilde z= \frac{1}{\sqrt2}(x^0+ix^1),~
w=-\frac{1}{\sqrt2}(x^2-ix^3),~ 
\widetilde w= \frac{1}{\sqrt2}(x^2+ix^3),
\end{eqnarray}
which satisfy the Euclidean metric $ds^2=(dx^0)^2+(dx^1)^2+(dx^2)^2+(dx^3)^2$.
Then Eq.\eqref{asdym_cpx} reduces to the anti-self-dual 
Yang-Mills equation:
$F_{01}+F_{23}=0,~~F_{02}-F_{13}=0,~~F_{03}+F_{12}=0$. 

Further, the condition $\overline{M}=L$ gives rise to
the relations $\gamma=\overline{\lambda}\overline{\beta},~
\delta=-\overline{\lambda}\overline{\alpha}, 
~\mu=-1/\overline{\lambda}$,
and the soliton solution 
\eqref{soliton_cpx} could be represented by 
\begin{eqnarray}
\label{soliton_E}
Q=\left(
\begin{array}{cc}
a_1e^{L}+a_2e^{-L}
& 
b_1e^{\overline{L}}+b_2e^{-\overline{L}}
\\ 
-\overline{b}_1e^{L}-\overline{b}_2e^{-L}
& 
\overline{a}_1e^{\overline{L}}
+\overline{a}_2e^{-\overline{L}} 
\end{array}\right),~~~
\Lambda=
\left(
\begin{array}{cc}
\lambda & 0 \\
0 & -1/\overline{\lambda}
\end{array}
\right),
\end{eqnarray}
where $L=(\lambda \beta) z
+\alpha\overline{z}
+(\lambda \alpha) w
-\beta \overline{w}$.
The real coordinates expansion of it is
\begin{eqnarray}
 L=l_\mu x^{\mu}, ~~~
l_\mu=\frac{1}{\sqrt{2}}\left(\alpha+\lambda\beta,~
i(\alpha-\lambda\beta),~\beta-\lambda\alpha,~i(\beta+\lambda\alpha)
\right).
\end{eqnarray}
Under these setting, 
the action density of the soliton solution \eqref{soliton_E} is
\begin{eqnarray}
\label{action_E}
{\mbox{Tr}} F_{\mu \nu}F^{\mu \nu}\!=\!
8\left[(\left|\alpha\right|^2\!+\left|\beta\right|^2)
(\left|\lambda \right|^2\!+1) \left|\varepsilon_0 \right| \right]^2\!
\left[\frac{2\varepsilon_1\widetilde{\varepsilon}_1
	\sinh^2 X_1
	-2\left|\varepsilon_2 \right|^2
	\sinh^2 X_2 
	-\left|\varepsilon_0 \right|^2}
{\left((\varepsilon_1\widetilde{\varepsilon}_1)^{\frac12}
	\cosh X_1 
	+\left|\varepsilon_2 \right|
	\cosh X_2\right)^4}
\right]\!\!,
\end{eqnarray}
where
\begin{eqnarray}
\label{X}
&&X_1=\overline {L} + L+\dfrac12 \log(\varepsilon_1/\widetilde{\varepsilon}_1),~
X_2=\overline{L} - L+\dfrac12 \log(\varepsilon_2/\overline{\varepsilon}_2),\\
&&\varepsilon_0=a_1 \overline{b}_2-a_2\overline{b}_1,~ \\
&& \varepsilon_1=\left|a_1\right|^2 + \left|b_1 \right|^2,~
\widetilde{\varepsilon}_1=\left|{a}_2 \right|^2 + \left|{b_2} \right|^2 \in \mathbb{R},\\
&&\varepsilon_2=\overline{a}_1a_2+b_1\overline{b}_2.
\label{varepsilon_2}
\end{eqnarray}
Note that the action density
vanishes identically when $\alpha=\beta=0$ or $\varepsilon_0=0$. 

To realize the gauge group to be $G=U(N)$, the action density Tr$F_{\mu\nu}F^{\mu\nu}$
should be negative definite  
because $F_{\mu\nu}$ is anti-hermitian and 
eigenvalues of it are pure imaginary. 
However, the action density \eqref{action_E}
is not negative definite at any point on $\mathbb{E}$.
This implies that the gauge group cannot be unitary. 

However, action density Tr$F_{\mu\nu}F^{\mu\nu}$ could be real-valued even though gauge group is not unitary. 
Note that this configuration has solitonic behavior in the 
the $X_1$-direction and periodic behavior in the $X_2$-direction
because $X_1$ is clearly real and
$X_2$ is pure-imaginary, implying  
$\cosh X_2= \cos \left({\mbox{Im}} X_2\right), 
\sinh X_2 = i\sin \left({\mbox{Im}} X_2\right)$. 
By this property and \eqref{X}$\sim$\eqref{varepsilon_2}, 
Tr$F_{\mu\nu}F^{\mu\nu}$ is clearly real-valued.
Furthermore, since $\cosh X_1\geq 1$, $-1\leq \cosh X_2=\cos 
\left({\mbox{Im}} X_2\right)\leq 1$ 
and 
$\varepsilon_1\widetilde{\varepsilon}_1\geq \vert \varepsilon_2\vert^2$, 
all singularities appear on the locus
$D:=\left\{x^\mu\in\mathbb{R}^4~|~\cosh X_1= 1,~ 
\cosh X_2=\cos \left({\mbox{Im}} X_2\right)= -1,~
\varepsilon_1\widetilde{\varepsilon}_1=\vert \varepsilon_2\vert^2
\right\}$. 
As we mentioned in section 2.2 below \eqref{epsilon2}, the final condition 
$\varepsilon_1\widetilde{\varepsilon}_1=\vert \varepsilon_2\vert^2$
in $D$ implies 
$\mbox{Tr}F_{\mu\nu}F^{\mu\nu}=0$.
Therefore, there is no singularity in the action density. The same discussion is also valid for other signatures.

Another surprising thing comes when we focus only on solitonic behavior part 
by setting $a_2=b_1=0$ in \eqref{action_E}. 
Then the soliton solution 
\begin{eqnarray}
\label{1soliton_E}
Q=\left(
\begin{array}{cc}
ae^{L}
& 
be^{-\overline{L}}
\\ 
-\overline{b}e^{-L}
& 
\overline{a}e^{-\overline{L}} 
\end{array}\right),
\end{eqnarray}
leads to a simpler form of action density:
\begin{eqnarray}
\label{1action_E}
{\mbox{Tr}} F_{\mu \nu}F^{\mu \nu}=
8\left[(\left|\alpha\right|^2+\left|\beta\right|^2)
(\left|\lambda \right|^2+1) \ \right]^2
\left(
2{\mbox{sech}}^2 X-3{\mbox{sech}}^4 X
\right),
\end{eqnarray}
where $X=L +\overline{L} +\log(\left|a \right| /\left|b \right|)
$. 
(Note that $\left| \varepsilon_0 \right|^2=
\varepsilon_1 \widetilde{\varepsilon}_1 = \left| ab \right|^2,~
\left| \varepsilon_2 \right|^2=0$.)

We find that the action density has its principal peak on 
a three-dimensional hyperplane
defined by $X=L+ \overline{L}+\log(\left|a \right| /\left|b \right|)=0$
with normal vector $l_\mu+\overline{l}_\mu$.  
Therefore, it's a domain wall in $\mathbb{R}^4$. 
More surprisingly, 
integration of this action density over $\mathbb{E}$ is 
suggested to be 
zero. 
In order to explain this property, 
let us introduce three independent axes $X^1, X^2, X^3$
in the directions orthogonal to the $X$-axis 
(normal direction of the domain wall (DW)). 
Then, integration of the action density 
would be performed naively by the following finite box regularization:
\begin{eqnarray}
\label{integ}
\int_{\mathbb{E}}{\mbox{Tr}} F_{\mu \nu}F^{\mu \nu}d^4x
&\propto& \lim_{R\rightarrow\infty}
\int_{-R}^{R} 
\int_{-R}^{R} 
\int_{-R}^{R} 
 dX^1dX^2dX^3
\int_{-R}^{R} 
(2{\mbox{sech}}^2 X-3{\mbox{sech}}^4 X) dX\nonumber\\
&=&
\int_{\scriptsize{\mbox{DW}}} dX^1dX^2dX^3
\int_{-\infty}^{\infty} 
(2{\mbox{sech}}^2 X-3{\mbox{sech}}^4 X) dX \nonumber\\.
&=& \int_{\scriptsize{\mbox{DW}}} dX^1dX^2dX^3 \left.({\mbox{tanh}}X \cdot {\mbox{sech}^2X}) \right|_{-\infty}^{~\infty} = 0.
\end{eqnarray}
This result suggests
that the soliton solution \eqref{1soliton_E} 
belongs to the sector of instanton number zero.  
The same discussion is also valid for other signatures.
We note that the present discussion lacks mathematical rigor. 
In order to justify the integration, 
we have to solve the anti-self-dual Yang-Mills equation 
originally with a suitable boundary condition compatible to the box 
regularization.\footnote{The authors thank an 
anonymous referee to point this out.}
This issue will be reported elsewhere.


Finally, we remark that
$J \in U(2) \Leftrightarrow \displaystyle  
	\Lambda=
	\left(
	\begin{array}{cc}
		\ e^{i\theta} & 0 \\
		0 & -e^{i\theta}
	\end{array}
	\right)$ 
and 
      $J \in SU(2) \Leftrightarrow  \displaystyle 
      \Lambda = \pm
      \left(
      \begin{array}{cc}
      \ i & 0 \\
      0 & -i
      \end{array}
      \right)$ $(\theta\in \mathbb{R})$
on $\mathbb{E}$.

In this section, we call 
solutions like \eqref{soliton_E}
one-soliton solutions and 
solutions like \eqref{1soliton_E}
pure one-soliton solutions to distinguish them.

\subsection{On Minkowski Real Space $\mathbb{M}$}

As discussed in the Euclidean case,  
we can take the following combination of real coordinates
$x^0,x^1,x^2,x^3$ on $\mathbb{M}$ to realize
the real slice condition: 
$z, \widetilde z \in \mathbb{R},~\widetilde w= \overline w$
\begin{eqnarray}
z=\frac{1}{\sqrt2}(x^0-x^1),~ 
\widetilde z= \frac{1}{\sqrt2}(x^0+x^1),~
w=\frac{1}{\sqrt2}(x^2-ix^3),~ 
\widetilde w= \frac{1}{\sqrt2}(x^2+ix^3),
\end{eqnarray}
which satisfy the Minkowski metric
$ds^2=(dx^0)^2-(dx^1)^2-(dx^2)^2-(dx^3)^2$.
Then Eq.\eqref{asdym_cpx} reduces to the anti-self-dual  
Yang-Mills equation:
$F_{01}+iF_{23}=0,~~F_{02}-iF_{13}=0,~~F_{03}+iF_{12}=0$. 
Due to the ASD equation, the realization of gauge group $G=U(N)$ 
is impossible since gauge fields 
$A_0, A_1, A_2$ and $A_3$ could not be all anti-hermitian. 

Further, the condition $\overline{M}=L$ yields 
relations $\beta=\overline{\mu}\alpha,~
\gamma=\overline{\alpha},~\delta=\overline{\lambda}\overline{\alpha}
$ (Relation between $\lambda$ and $\mu$ is not necessary.)
and the one-soliton 
solution \eqref{soliton_cpx} could be represented by 
\begin{eqnarray}
\label{solution_M}
Q&=&\left(
\begin{array}{cc}
a_1e^{L}+a_2e^{-L}
& 
b_1e^{\overline{L}}+b_2e^{-\overline{L}}
\\ 
-\overline{b}_1e^{L}-\overline{b}_2^{-L}
& 
\overline{a}_1e^{\overline{L}}
+\overline{a}_2e^{-\overline{L}} 
\end{array}\right),~~~
\Lambda=
\left(
\begin{array}{cc}
\lambda & 0 \\
0 & \mu
\end{array}
\right),\\
L&=&(\lambda \overline{\mu}\alpha) z
+\alpha\widetilde{z}
+(\lambda \alpha) w
+(\overline{\mu}\alpha)\overline{w}\\
&=& l_\mu x^\mu,~~~
l_\mu=\frac{1}{\sqrt{2}}\left( (1+\lambda \overline{\mu})\alpha,~
(1-\lambda \overline{\mu})\alpha,~
(\overline{\mu}+\lambda)\alpha,~
i(\overline{\mu}-\lambda)\alpha
\right).
\label{dispersion_M}
\end{eqnarray}
Under these setting, the action density of the solution \eqref{solution_M} is 
\begin{eqnarray}
{\mbox{Tr}} F_{\mu \nu}F^{\mu \nu}=
8\left|{\alpha}(\lambda-\mu)\right|^4
\left|\varepsilon_0\right|^2
\left[\frac{2\varepsilon_1\widetilde{\varepsilon}_1
	\sinh^2 X_1
	-2\left|\varepsilon_2 \right|^2
	\sinh^2 X_2 
	-\left|\varepsilon_0 \right|^2}
{\left((\varepsilon_1\widetilde{\varepsilon}_1)^{\frac12}
	\cosh X_1 
	+\left|\varepsilon_2 \right|
	\cosh X_2\right)^4}
\right],
\end{eqnarray}
where
\begin{eqnarray}
&&X_1=\overline{L} + L+\dfrac12 \log(\varepsilon_1/\widetilde{\varepsilon}_1),~
X_2=\overline{L} - L+\dfrac12 \log(\varepsilon_2/\overline{\varepsilon}_2),\\
&&\varepsilon_0=a_1 \overline{b}_2-a_2\overline{b}_1,~ \\
&& \varepsilon_1=\left|a_1\right|^2 + \left|b_1 \right|^2,~
\widetilde{\varepsilon}_1=\left|{a}_2 \right|^2 + \left|{b_2} \right|^2 
\in \mathbb{R},\\
&&\varepsilon_2=\overline{a}_1a_2+b_1\overline{b}_2.
\end{eqnarray}
Note that the action density
vanishes identically when $\lambda=\mu$ or 
$\alpha=0$ or $\varepsilon_0=0$. 

The pure one-soliton solution 
is given by the same trick as in the Euclidean case:
\begin{eqnarray}
\label{1soliton_M}
Q=\left(
\begin{array}{cc}
ae^{L}
& 
be^{-\overline{L}}
\\ 
-\overline{b}e^{-L}
& 
\overline{a}e^{-\overline{L}} 
\end{array}\right),
\end{eqnarray}
which leads to the action density
\begin{eqnarray}
\label{1action_M}
{\mbox{Tr}} F_{\mu \nu}F^{\mu \nu}=
8\left|{\alpha}^2(\lambda-\mu)\right|^2
\left(
2{\mbox{sech}}^2 X-3{\mbox{sech}}^4 X
\right),
\end{eqnarray}
where 
$X=L + \overline{L} +\log(\left|a \right| /\left|b \right| )$. 
(Note that 
$\left| \varepsilon_0 \right|^2=
\varepsilon_1 \widetilde{\varepsilon}_1 = \left| ab \right|^2,~
\left| \varepsilon_2 \right|^2=0$.)
Once again, integration of the action density \eqref{1action_M} 
over $\mathbb{M}$
vanishes by the same reason as in \eqref{integ}. 

Finally, we remark that 
	$J \in U(2)\Leftrightarrow
	\Lambda=
	\left(
	\begin{array}{cc}
	\ e^{i\theta_1} & 0 \\
	0 & e^{i\theta_2}
	\end{array}
	\right)$ and 
	$J \in SU(2) \Leftrightarrow 
	\Lambda =
	\left(
	\begin{array}{cc}
	\ e^{i\theta} & 0 \\
	0 & e^{-i\theta}
	\end{array}
	\right)$  $(\theta_1, \theta_2\in \mathbb{R})$
on $\mathbb{M}$.

\subsection{On Ultrahyperbolic Real Space $\mathbb{U}_1$}

The discussion of $\mathbb{U}_1$ is quite similar to the Euclidean case.   
We can take the following combination of the real coordinates
$x^0,x^1,x^2,x^3$ on $\mathbb{U}_1$ to realize
the real slice condition:
$\widetilde z=\overline z,~\widetilde w= \overline w$
\begin{eqnarray}
z=\frac{1}{\sqrt2}(x^0-ix^1),~ 
\widetilde z= \frac{1}{\sqrt2}(x^0+ix^1),~
w=\frac{1}{\sqrt2}(x^2-ix^3),~ 
\widetilde w= \frac{1}{\sqrt2}(x^2+ix^3),~ 
\end{eqnarray}
which satisfy the Ultrahyperbolic metric 
$ds^2=(dx^0)^2+(dx^1)^2-(dx^2)^2-(dx^3)^2$. 
Then Eq.\eqref{asdym_cpx} reduces to the anti-self-dual 
Yang-Mills equation:
$F_{01}-F_{23}=0,~~F_{02}-F_{13}=0,~~F_{03}+F_{12}=0$. 

Further, the condition $\overline{M}=L$ yields  
the relations $\gamma=\overline{\lambda}\overline{\beta},~
\delta=\overline{\lambda}\overline{\alpha}, 
~\mu=1/\overline{\lambda}$, 
and the one-soliton solution 
\eqref{soliton_cpx} could be represented by 
\begin{eqnarray}
\label{soliton_U1}
Q&=&\left(
\begin{array}{cc}
a_1e^{L}+a_2e^{-L}
& 
b_1e^{\overline{L}}+b_2e^{-\overline{L}}
\\ 
-\overline{b}_1e^{L}-\overline{b}_2e^{-L}
& 
\overline{a}_1e^{\overline{L}}
+\overline{a}_2e^{-\overline{L}} 
\end{array}\right),~~~
\Lambda=
\left(
\begin{array}{cc}
\lambda & 0 \\
0 & 1/\overline{\lambda}
\end{array}
\right),\\
L&=&(\lambda \beta) z
+\alpha\overline{z}
+(\lambda \alpha) w
+\beta \overline{w}\\
&=&l_\mu x^{\mu}, ~~~
l_\mu=\frac{1}{\sqrt{2}}\left(\alpha+\lambda\beta,~
i(\alpha-\lambda\beta),~\beta+\lambda\alpha,~i(\beta-\lambda\alpha)
\right).
\end{eqnarray}
Under these setting, the action density of the solution \eqref{soliton_U1} is
\begin{eqnarray}
\label{action_U1}
{\mbox{Tr}} F_{\mu \nu}F^{\mu \nu}\!=\!
8\left[(\left|\alpha\right|^2 \!-\! \left|\beta\right|^2)
(\left|\lambda \right|^2 \!-\! 1) 
\vert \varepsilon_0\vert \right]^2\!\!
\left[\frac{2\varepsilon_1\widetilde{\varepsilon}_1
	\sinh^2 X_1
	-2\left|\varepsilon_2 \right|^2
	\sinh^2 X_2 
	-\left|\varepsilon_0 \right|^2}
{\left((\varepsilon_1\widetilde{\varepsilon}_1)^{\frac12}
	\cosh X_1 
	+\left|\varepsilon_2 \right|
	\cosh X_2\right)^4}
\right]\!\!,
\end{eqnarray}
where
\begin{eqnarray}
&&X_1=\overline L + L+\dfrac12 \log(\varepsilon_1/\widetilde{\varepsilon}_1),~
X_2=\overline L - L+\dfrac12 \log(\varepsilon_2/\overline{\varepsilon}_2),\\
&&\varepsilon_0=a_1 \overline{b}_2-a_2\overline{b}_1,~ \\
&& \varepsilon_1=\left|a_1\right|^2 + \left|b_1 \right|^2,~
\widetilde{\varepsilon}_1=\left|{a}_2 \right|^2 + \left|{b_2} \right|^2 \in \mathbb{R},\\
&&\varepsilon_2=\overline{a}_1a_2+b_1\overline{b}_2.
\end{eqnarray}
Note that the action density
vanishes identically when $\vert\alpha\vert=\vert\beta\vert$ 
or $\vert\lambda\vert=1$ or $\varepsilon_0=0$. 

The pure one-soliton solution is given by
\begin{eqnarray}
\label{1soliton_U1}
Q=\left(
\begin{array}{cc}
ae^{L}
& 
be^{-\overline{L}}
\\ 
-\overline{b}e^{-L}
& 
\overline{a}e^{-\overline{L}} 
\end{array}\right),
\end{eqnarray}
which leads to the action density 
\begin{eqnarray}
\label{1action_U1}
{\mbox{Tr}} F_{\mu \nu}F^{\mu \nu}=
8\left[(\left|\alpha\right|^2-\left|\beta\right|^2)
(\left|\lambda \right|^2-1) \right]^2
\left(
2{\mbox{sech}}^2 X-3{\mbox{sech}}^4 X
\right),
\end{eqnarray}
where $X=L + \overline{L} +\log(\left|a \right| /\left|b \right| )$.
(Note that 
$\left| \varepsilon_0 \right|^2=
\varepsilon_1 \widetilde{\varepsilon}_1 = \left| ab \right|^2,~~
\left| \varepsilon_2 \right|^2=0$.)
Integration of the action density \eqref{1action_U1} over $\mathbb{U}_1$ 
is zero again by the same reason as in \eqref{integ}. 

Finally, we remark that the condition 
$J \in U(2)$ implies Tr$F_{\mu \nu} F^{\mu \nu} =0$ on $\mathbb{U}_1$. 
	In fact, the gauge group can not be unitary
	under the gauge condition $A_{\widetilde{z}}=A_{\widetilde{w}}=0$
	on $\mathbb{U}_{1}$ as well because
	$\sqrt{2} A_{\widetilde{z}}=A_0-iA_1,  
	\sqrt{2}A_{\widetilde{w}}=A_2-iA_3$ implies $A_0=iA_1, A_2=iA_3$.
	Hence under this gauge, only one possible solution is $A_\mu=0$ for $G=U(N)$, that is, $F_{\mu\nu}=0$. 
	The vanishing field strength leads to the trivial action density Tr$F_{\mu\nu}F^{\mu\nu}=0$ which is valid in arbitrary gauge. 
	Therefore there is no $G=U(N)$ ASD gauge fields which give non-trivial action density.

\subsection{On Ultrahyperbolic Real Space $\mathbb{U}_2$}

Finally, we discuss another real slice of 
the Ultrahyperbolic signature, say $\mathbb{U}_2$. 
We take the following combination of real coordinates
$x^0,x^1,x^2,x^3$ on $\mathbb{U}_2$ to realize 
the real slice condition:$z, \widetilde z, w, \widetilde w\in \mathbb{R}$
\begin{eqnarray}
z=\frac{1}{\sqrt2}(x^0-x^2),~ 
\widetilde z= \frac{1}{\sqrt2}(x^0+x^2),~
w=-\frac{1}{\sqrt2}(x^1-x^3),~ 
\widetilde w= \frac{1}{\sqrt2}(x^1+x^3),~ 
\end{eqnarray}
which satisfy the Ultrahyperbolic signature $ds^2=(dx^0)^2+(dx^1)^2-(dx^2)^2-(dx^3)^2$. 
Then Eq.\eqref{asdym_cpx} reduces to the anti-self-dual  
Yang-Mills equation:
$F_{01}+F_{23}=0,~~F_{02}+F_{13}=0,~~F_{03}-F_{12}=0$. 

Further, the one-soliton solution 
\eqref{soliton_cpx} is reduced by the condition $\overline{M}=L$ 
($\Rightarrow \gamma=\overline{\alpha},~
\delta=\overline{\beta},~\mu=\overline{\lambda}$)
to the following:
\begin{eqnarray}
\label{soliton_U2}
Q&=&\left(
\begin{array}{cc}
a_1e^{L}+a_2e^{-L}
& 
b_1e^{\overline{L}}+b_2e^{-\overline{L}}
\\ 
-\overline{b}_1e^{L}-\overline{b}_2e^{-L}
& 
\overline{a}_1e^{\overline{L}}
+\overline{a}_2e^{-\overline{L}} 
\end{array}\right),~~~
\Lambda=
\left(
\begin{array}{cc}
\lambda & 0 \\
0 & \overline{\lambda}
\end{array}
\right),\\
L&=&(\lambda \beta) z
+\alpha\widetilde{z}
+(\lambda \alpha) w
+\beta \widetilde{w},\\
&=&l_\mu x^{\mu}, ~~~
l_\mu=\frac{1}{\sqrt{2}}\left(\alpha+\lambda\beta,~
\beta-\lambda\alpha,~\alpha-\lambda\beta,~\beta+\lambda\alpha),
\right.
\end{eqnarray}
which leads to the action density
\begin{eqnarray}
\label{action_U2}
{\mbox{Tr}} F_{\mu \nu}F^{\mu \nu}\!=\!
8\left[(\alpha\overline{\beta} \!-\! \overline{\alpha}\beta)
(\lambda \!-\!\overline{\lambda}) \left|\varepsilon_0 \right| \right]^2
\!
\left[\frac{2\varepsilon_1\widetilde{\varepsilon}_1
	\sinh^2 X_1
	-2\left|\varepsilon_2 \right|^2
	\sinh^2 X_2 
	-\left|\varepsilon_0 \right|^2}
{\left((\varepsilon_1\widetilde{\varepsilon}_1)^{\frac12}
	\cosh X_1 
	+\left|\varepsilon_2 \right|
	\cosh X_2\right)^4}
\right]\!\!,	
\end{eqnarray}
where
\begin{eqnarray}
&&X_1=\overline{L} + L+\dfrac12 \log(\varepsilon_1/\widetilde{\varepsilon}_1),~
X_2=\overline{L} - L+\dfrac12 \log(\varepsilon_2/\overline{\varepsilon}_2),\\
&&\varepsilon_0=a_1 \overline{b}_2-a_2\overline{b}_1,~ \\
&& \varepsilon_1=\left|a_1\right|^2 + \left|b_1 \right|^2,~
\widetilde{\varepsilon}_1=\left|{a}_2 \right|^2 + \left|{b_2} \right|^2 \in \mathbb{R},\\
&&\varepsilon_2=\overline{a}_1a_2+b_1\overline{b}_2.
\end{eqnarray}
Note that the action density
vanishes identically when $\alpha\overline{\beta}
\in \mathbb{R}$ or 
$\lambda\in \mathbb{R}$ or
$\varepsilon_0=0$. 

The pure one-soliton solution is given by 
\begin{eqnarray}
\label{1soliton_U2}
Q=\left(
\begin{array}{cc}
ae^{L}
& 
be^{-\overline{L}}
\\ 
-\overline{b}e^{-L}
& 
\overline{a}e^{-\overline{L}} 
\end{array}\right),
\end{eqnarray}
and the action density becomes
\begin{eqnarray}
\label{1action_U2}
{\mbox{Tr}} F_{\mu \nu}F^{\mu \nu}=
8\left[(\alpha\overline{\beta}-\overline{\alpha}\beta)
(\lambda -\overline{\lambda})  \right]^2
\left(2{\mbox{sech}}^2 X-3{\mbox{sech}}^4 X\right),
\end{eqnarray}
where $X=L + \overline{L} +\log(\left|a \right| /\left|b \right|)$.
(Note that 
$\left| \varepsilon_0 \right|^2=
\varepsilon_1 \widetilde{\varepsilon}_1 = \left| ab \right|^2,~
\left| \varepsilon_2 \right|^2=0$.)
By the same reason as in \eqref{integ}, 
integration of the action density \eqref{1action_U2} 
over $\mathbb{U}_2$ is zero.

Finally, we remark that $J \in U(2) \Leftrightarrow   J \in SU(2) \Leftrightarrow
	\Lambda=
	\left(
	\begin{array}{cc}
	\ e^{i\theta} & 0 \\
	0 & e^{-i\theta}
	\end{array}
	\right)$
on $\mathbb{U}_2$.
In fact, we even find that the gauge group can be unitary in this case ! 
First of all, gauge fields $A_z$ and $A_w$ are anti-hermitian 
on $\mathbb{U}_2$ naturally (See \eqref{A_k_2}, \eqref{A_k_3}).
On the other hand, $\sqrt{2}A_z=A_0+A_2,\sqrt{2}A_{\widetilde{z}}=A_0-A_2, 
\sqrt{2}A_w=A_1+A_3, \sqrt{2}A_{\widetilde{w}}=A_1-A_3$
together with $A_{\widetilde{z}}=A_{\widetilde{w}}=0$ implies all gauge fields $A_\mu$ must be anti-hermitian.
That is, $G=SU(2)$ gauge theory is realized on $\mathbb{U}_2$ successfully.

\section{Comparison to already known soliton solutions}

In this section, we review already known soliton solutions
of the anti-self-dual Yang-Mills equation. 
The four-dimensional complex coordinates 
$(z,\widetilde z,w,\widetilde w)$ used here is defined as in section 2.1.

\subsection{Atiyah-Ward ansatz solutions ($G=GL(2)$)}

Firstly, we begin with the Atiyah-Ward ansatz solutions 
\cite{AtWa}. The simplest one is \cite{CFGY}:
\begin{eqnarray}
J =
\left(
\begin{array}{cc}
0 & -1 \\
1 & \Delta_{0}
\end{array}
\right)  
\end{eqnarray}
with a scalar function $\Delta_0(x)$ 
and the Yang equation reduces to a simpler linear equation
\begin{eqnarray}
(\partial_{\widetilde z}\partial_{z}-\partial_{\widetilde w} \partial_{w})\Delta_{0}=0.
\end{eqnarray}
A natural one-soliton solution is given by
\begin{eqnarray}
\label{natural1}
\Delta_{0}= \frac{1}{2}(e^{L}+e^{-L}) = {\mbox{cosh}}L,
~~~ L=(\lambda\beta) z + \alpha \widetilde z 
 + (\lambda \alpha) w + \beta \widetilde w,
\end{eqnarray}
and the corresponding action density vanishes: Tr$F^2=0$ by simple calculation.

The second simplest one is given in the following ansatz:
\begin{eqnarray}
J=
\left(
\begin{array}{cc}
\Delta_0-\Delta_1 \Delta_0^{-1} \Delta_{-1} & -\Delta_1 \Delta_0^{-1} \\
\Delta_0^{-1} \Delta_{-1} & \Delta_0^{-1} 
\end{array}
\right),
\end{eqnarray}
which relates to Yang's R-gauge \cite{Yang} and include the non-linear plane wave solutions 
in the Minkowski signature \cite{deVega}. 
By substituting $J$-matrix into the Yang equation,
it reduces to the following chasing equations
\begin{eqnarray}
\partial_{z} ~\Delta_{i} = \partial_{\widetilde w} ~\Delta_{i+1},~~
\partial_{w} ~\Delta_{i} = \partial_{\widetilde z} ~\Delta_{i+1},
\end{eqnarray}
which implies that $\Delta_0$ solves the Laplace equation: 
$(\partial_{\widetilde z}\partial_{z}-\partial_{\widetilde w} 
\partial_{w})\Delta_{0}=0$.

A natural one-soliton solution is given by
\begin{eqnarray}
\label{natural2}
\Delta_{0}= {\mbox{cosh}}L,~~
\Delta_{1}= \lambda {\mbox{cosh}}L,~~
\Delta_{-1}= \lambda^{-1} {\mbox{cosh}}L,
\end{eqnarray}
and the corresponding action density is trivial again: Tr$F^2=0$ by simple calculation. 

\subsection{'t Hooft ansatz solutions ($G=SU(2)$)}

The 't Hooft ansatz \cite{tHooft} 
(or known as the Corrigan-Fairlie-'t Hooft-Wilczek ansatz \cite{CoFa,Wilczek})
is very important for the study of 
$G=SU(2)$ gauge theory on the four-dimensional 
Euclidean space and is given by 
\begin{eqnarray}
 A_{\mu} = i\eta_{\mu\nu}^{(+)a}\sigma_a \partial^{\nu}\log \varphi,
\end{eqnarray}
where $\eta_{\mu\nu}^{(+)a}~(a=1,2,3)$ is the self-dual 
't Hooft symbol and $\sigma_a$ is the Pauli matrices.
Under the 't Hooft ansatz,
the anti-self-dual Yang-Mills equation reduces to the Laplace equation
\begin{eqnarray}
\label{laplace}
(\partial_{\overline z}\partial_{z}+\partial_{\overline w} 
\partial_{w})\varphi=0.
\end{eqnarray}
A natural one-soliton solution is given by
\begin{eqnarray}
\label{natural3}
\varphi= \frac{1}{2}(e^{K}+e^{-K}) = {\mbox{cosh}}K,~~~
K:=k_\mu x^\mu,
\end{eqnarray}
where $k_\mu$ are real constants
which satisfy $k^2=k_\mu k^\mu=0$ due to \eqref{laplace}.
By using some formulas on the 't Hooft symbol,
we can easily show that
\begin{eqnarray}
{\mbox{Tr}}F^2 = -3 (k^2)^2 (4{\mbox{sech}}^4K -5{\mbox{sech}}^2K+2)
\stackrel{k^2=0}{=}0.
\end{eqnarray}

In conclusion, the action density of the natural 
one-soliton solutions \eqref{natural1},  \eqref{natural2} and 
\eqref{natural3}  are all trivial.

\section{Conclusion and Discussion}

In this paper, we constructed exact soliton solutions
of four-dimensional anti-self-dual Yang-Mills equations for $G = GL(2)$ which possess real-valued action densities. 
Our results showed that such type of solitons can be interpreted as domain wall 
in four-dimensional spaces and
 $G=U(2)$ solitons exist on the Ultrahyperbolic signature $\mathbb{U}_2$. 
This fact has a strong connection with N=2 string theories \cite{ Marcus,OoVa}.

In N=2 string theories,
	the equation of motion of the effective action is the Ultrahyperbolic space $\mathbb{U}_2$ version of anti-self-dual Yang-Mills equation for $G=U(2)$. Therefore,
our soliton solutions obtained in section 3.4 might
give us a hint for finding the corresponding physical objects 
in these theories. 
On the other hand, the Euclidean and Minkowski signature version of
such kind of $G=U(2)$ domain wall solutions (or the non-abelian
plane waves \cite{Coleman_PLB} of the Yang-Mills equation)   
are still unknown and worth investigating for our future work.
These studies might perhaps relate to new perturbative aspects of quantum field
theories, new invariants in the four-dimensional geometry 
or the origin of dark matters someday.

For multi-soliton solutions, we presented these discussions on noncommutative Euclidean spaces explicitly by the noncommutative Darboux transformation in \cite{GHHN}
and the noncommutative B\"acklund transformation in \cite{GHN2, HaOk}.
Asymptotic behaviors of these noncommutative soliton solutions
were also proved to be the same as in commutative spaces \cite{GHHN, HaOk}.
It might be an interesting future work to confirm 
our conjecture that $n$ soliton solutions in \cite{GHHN, GHN2, HaOk} 
have $n$ isolated localized lumps of energy and
preserve their shapes and velocities 
on each localized solitary wave lump.
In addition, explicit analysis of these $n$ soliton scatterings would be expected to give the phase shifts 
in the scattering processes as discussed 
in the standard soliton theory (e.g. \cite{MaSa}).  

Another interesting problem is to compare the asymptotic behaviors of 
our solutions \cite{GHHN, GHN2, HaOk} with
multi-soliton solutions in \cite{deVega}.
All of these studies  might lead to a general formulation of
Kodama's Grassmannian approach to the study of soliton scatterings 
\cite{Kodama},  
and give a new insight into Hirota's bilinear forms \cite{MOS}
 or other formulation of integrable hierarchies \cite{Nakamura, Takasaki_CMP}.

\subsection*{Acknowledgments}

MH is grateful to Jon Nimmo for fruitful discussion and warm hospitality
when he visited Glasgow especially in 2006 and 2008.
MH also thanks Kanehisa Takasaki for discussion.
The work of MH is supported
by Grant-in-Aid for Scientific Research (\#16K05318).
SCH would like to thank Jon Nimmo and Claire Gilson 
for private notes and Peiqiang Lin for discussion. 
The work of SCH is supported by the scholarship 
of Japan-Taiwan Exchange Association.


\appendix

\section{Calculation of Action Density  \eqref{action_cpx}}

\subsection*{Yang's $J$-matrix}
\begin{eqnarray}
\label{N=2_com}
J&=&
-Q\Lambda^{-1}Q^{-1}
=
\frac{-1}{\Delta}\left(
\begin{array}{cc}
\lambda^{-1} AD - \mu^{-1} BC
& 
(\mu^{-1}-\lambda^{-1})AB
\\ 
(\lambda^{-1} - \mu^{-1})CD
& 
\mu^{-1} AD - \lambda^{-1} BC
\end{array}\right),\\
Q&=&
\left(
\begin{array}{cc}
A & B
\\ 
C & D
\end{array}\right),~
\Delta:=\det Q=AD-BC
\end{eqnarray}

\subsection*{Derivative of $J$-matrix}

\begin{eqnarray}
\label{N=2_com}
\begin{array}{ll}
\displaystyle
J^\prime=\frac{\mu^{-1}-\lambda^{-1}}{\Delta^2}
\left(
\begin{array}{cc}
E
& 
F
\\ 
G
& 
-E
\end{array}\right)
~~~~
& 
\left\{
\begin{array}{l}
E=(AC^\prime-A^\prime C)BD-(BD^\prime-B^\prime D)AC 
\\
F=-(AC^\prime-A^\prime C)B^2+(BD^\prime-B^\prime D)A^2 
\\
G=(AC^\prime-A^\prime C)D^2-(BD^\prime-B^\prime D)C^2 
\\
\end{array}
\right.
\end{array}
\end{eqnarray}

\subsection*{Gauge Field ($f^{\prime}:=\partial_
	{k} f,~k=z,~w$)}
\begin{eqnarray}
\label{A_k}
{A_{k}}&=&J^{-1} J^{\prime} =\frac{1}{\Delta^2}\left(
\begin{array}{cc}
R
& 
S
\\ 
T
& 
-R
\end{array}\right)
\\
&&
\left\{
\begin{array}{l}
R
=(\mu/\lambda-1)(AC^{\prime}-A^{\prime}C)BD
-(1-\lambda/\mu)(BD^{\prime}-B^{\prime}D)AC \notag \\
S
=-(\mu/\lambda-1)(AC^{\prime}-A^{\prime}C)B^2
+(1-\lambda/\mu)(BD^{\prime}-B^{\prime}D)A^2 \notag \\
T
=(\mu/\lambda-1)(AC^{\prime}-A^{\prime}C)D^2
-(1-\lambda/\mu)(BD^{\prime}-B^{\prime}D)C^2 \notag \\
\end{array}
\right.
\end{eqnarray}
Note that if we take $(Q,\Lambda)$ as mentioned in \eqref{soliton_cpx}, then a simple form  of $A_k$ would be found from the result~
$AC^{\prime}-A^{\prime}C=2\lambda p,~~
BD^{\prime}-B^{\prime}D=2\mu q$~:
\begin{eqnarray}
\label{A_k_2}
{A_{k}}&=&\frac{2(\mu-\lambda)}{\Delta^2}\left(
\begin{array}{cc}
pBD-qAC
& 
-pB^2+qA^2
\\ 
pD^2-qC^2
& 
-pBD+qAC
\end{array}\right)
\\
&&
\left\{
\begin{array}{l}
	(p,~q):=(\alpha\varepsilon_{0},~\gamma\widetilde{\varepsilon}_{0})~~
	{\mbox{if}} ~~ m=w,~~ 
	(p,~q):=(\beta\varepsilon_{0},~\delta\widetilde{\varepsilon}_{0})~~
	{\mbox{if}} ~~ m=z \notag \\
	\varepsilon_{0}:=a_{2}c_{1}-a_{1}c_{2},~~
	\widetilde{\varepsilon}_{0}:=b_{2}d_{1}-b_{1}d_{2}
\end{array}
\right.
\end{eqnarray}
Moreover, if we consider the Ultrahyperbolic signature $\mathbb{U}_{2}$ 
(Take $(Q,\Lambda)$ mentioned in \eqref{soliton_U2}), 
then gauge fields become anti-hermitian naturally:
\begin{eqnarray}
\label{A_k_3}
{A_{k}}&=&\frac{2(\overline{\lambda}-\lambda)}{\Delta^2}\left(
\begin{array}{cc}
p\overline{A}B+\overline{p}A\overline{B}
& 
-pB^2+\overline{p}A^2
\\ 
p\overline{A}^2-\overline{p}\overline{B}^2
& 
-p\overline{A}B-\overline{p}A\overline{B}
\end{array}\right)
\\
&&
\left\{
\begin{array}{l}
p:=\alpha\varepsilon_{0},~~{\mbox{if}} ~~ m=w,~~ 
p:=\beta\varepsilon_{0},~~{\mbox{if}} ~~ m=z,~~
\varepsilon_{0}:=a_{1}\overline{b}_{2}-\overline{a}_{2}b_{1},~~ \notag
\end{array}
\right.
\end{eqnarray}

\subsection*{Field Strength
($\dot{f}:=\partial_{l} f,~l=\widetilde z,~\widetilde w$ )}

\begin{eqnarray}
{F_{k l}}
&=&-\partial_{l}A_{k}
=\frac{2(\lambda-\mu)}{\Delta^2}\left(
\begin{array}{cc}
U
& 
V
\\ 
W
& 
-U
\end{array}\right)
\\
&&
\left\{
\begin{array}{l}
U
=p\left[\dot{B}D+B\dot{D}-2BD(\dot{\Delta}/\Delta)\right]
-q\left[\dot{A}C+A\dot{C}-2AC(\dot{\Delta}/\Delta)\right]  \notag \\
V
=-2p\left[B\dot{B}-B^{2}(\dot{\Delta}/\Delta)\right]
+2q\left[A\dot{A}-A^2(\dot{\Delta}/\Delta)\right]\notag \\
W
=2p\left[D\dot{D}-D^2(\dot{\Delta}/\Delta)\right]
-2q\left[C\dot{C}-C^2(\dot{\Delta}/\Delta)\right]\notag \\
\end{array}
\right.
\end{eqnarray}
Note that $p,q$ are defined as in \eqref{A_k_2} and $A_l=0$ as mentioned in \eqref{gauge_f_special}. 

\subsection*{Action density}

\begin{eqnarray}
\begin{split}
{\mbox{Tr}}{F_{w\widetilde{z}}F_{z\widetilde w}}
&=\frac{16(\lambda-\mu)^2 \varepsilon_{0}\widetilde{\varepsilon}_{0}}{\Delta^4}
\{4\varepsilon_{0}\widetilde{\varepsilon}_{0}\alpha \beta \gamma \delta + 
(\alpha \delta -\beta \gamma)^{2} (A\dot{D}-\dot{B}C) (\dot{A}D-B\dot{C}) 
\\
& ~~~~~~~~+\alpha \beta \gamma \delta [(AD-BC)
(\dot{A}\dot{D}
-\dot{B}\dot{C})
+(\dot{A}D-B\dot{C})
(A\dot{D}-\dot{B}C)]\},
\notag
\\
{\mbox{Tr}}{F^2_{w\widetilde{w}}}
&=\frac{16(\lambda-\mu)^2 \varepsilon_{0}\widetilde{\varepsilon}_{0}}{\Delta^4}
\{2\varepsilon_{0}\widetilde{\varepsilon}_{0}(\alpha^2\delta^2 +\beta^2 \gamma^2) + 
\\
& ~~~~~~~~+\alpha \beta \gamma \delta [(AD-BC)
(\dot{A}\dot{D}
-\dot{B}\dot{C})
+(\dot{A}D-B\dot{C})
(A\dot{D}-\dot{B}C)]\},
\notag \\
{\mbox{Tr}}F^2
&=
{\mbox{Tr}}{F_{mn}F^{mn}}
=4({\mbox{Tr}}{F_{w\widetilde{z}}F_{z\widetilde w}}-{\mbox{Tr}}{F^2_{w\widetilde{w}}})
\\
&=\frac{64(\lambda-\mu)^2(\alpha \delta -\beta \gamma)^2 \varepsilon_{0}\widetilde{\varepsilon}_{0}}{\Delta^4}
[(A\dot{D}-\dot{B}C) (\dot{A}D-B\dot{C})-2\varepsilon_{0}\widetilde{\varepsilon}_{0}] \\ 
\end{split}
\end{eqnarray}
Finally, substituting \eqref{soliton_cpx} into the above formula, we get 
\eqref{action_cpx}.




\begin{thebibliography}{99}
	
	
	\bibitem{Actor}
	A.~Actor,
	Rev.\ Mod.\ Phys.\  {\bf 51} (1979) 461.
	
	\bibitem{ADHM}
	M.~F.~Atiyah, N.~J.~Hitchin, V.~G.~Drinfeld and Y.~I.~Manin,
	Phys.\ Lett.\ A {\bf 65} (1978) 185.
	
	\bibitem{AtWa}
	M.~F.~Atiyah and R.~S.~Ward,  
	Commun.\ Math.\ Phys.\ {\bf 55} 117 (1977).
	
	\bibitem{Coleman_PLB}
	S.~Coleman,
	Phys.\ Lett.\ B {\bf 70}, 59 (1977).
	
	\bibitem{Coleman}
	S.~Coleman,
	{\it Aspects of Symmetry}
	(Cambridge UP, 1985).
	
	\bibitem{CoFa}
	E.~Corrigan and D.~B.~Fairlie,
	Phys.\ Lett.\ B {\bf 67}, 69 (1977).
	
	\bibitem{CFGY}
	E.~Corrigan, D.~B.~Fairlie, R.~G.~Yates and P.~Goddard,
	Commun.\ Math.\ Phys.\  {\bf 58}, 223 (1978).
	
	\bibitem{Monopole}
	N.~S.~Craigie, P.~Goddard and W.~Nahm,
	{\it Monopoles in Quantum Field Theory}
	(World Sci., 1982).
	
	\bibitem{deVega}
	H.~J.~de Vega,
	Commun.\ Math.\ Phys.\  {\bf 116}, 659 (1988).
	
	\bibitem{DHKM}
	N.~Dorey, T.~J.~Hollowood, V.~V.~Khoze and M.~P.~Mattis,
	Phys.\ Rept.\  {\bf 371} (2002) 231 
	[hep-th/0206063].
	
	\bibitem{GHHN}
	C.~R.~Gilson, M.~Hamanaka and S.C. Huang and J.~J.~C.~Nimmo,
        to appear in J.\ Phys.\ A
	[arXiv:2004.01718] \url{https://doi.org/10.1088/1751-8121/aba72e}.
	
	\bibitem{GHN2}
	C.~R.~Gilson, M.~Hamanaka and J.~J.~C.~Nimmo,
	Proc.\ Roy.\ Soc.\ Lond.\ A {\bf 465}, 2613 (2009)
	[arXiv:0812.1222].
	
	\bibitem{GPY}
	D.~J.~Gross, R.~D.~Pisarski and L.~G.~Yaffe,
	Rev.\ Mod.\ Phys.\ {\bf 53} (1981) 43.
	
	\bibitem{HaOk}
	M.~Hamanaka and H.~Okabe,
	Theor.\ Math.\ Phys.\  {\bf 197} (2018) 1451 
	[Teor.\ Mat.\ Fiz.\  {\bf 197} (2018) 68]
	[arXiv:1806.05188].
	
	\bibitem{Kodama}
	Y.~Kodama,
	{\it KP Solitons and the Grassmannians},
	(Springer, 2017). 
	
	\bibitem{MaSu}
	N.~S.~Manton and P.~Sutcliffe,
	{\it Topological solitons},
	(Cambridge UP, 2004).
	
	\bibitem{Marcus}
	N.~Marcus,
	Nucl.\ Phys.\ B {\bf 387} (1992) 263
	[hep-th/9207024].
	
	\bibitem{MaWo}
	L.~J.~Mason and N.~M.~Woodhouse,
	{\it Integrability, Self-Duality, and Twistor Theory}
	(Oxford UP, 1996). 
	
	\bibitem{MaSa}
	V.~B.~Matveev and M.~A.~Salle,
	{\it Darboux Transformations and Solitons},
	(Springer-Verlag, 1991).
	
	\bibitem{Nakamura}
	Y.~Nakamura,
	J.\ Math.\ Phys.\ {\bf 29}, 244 (1988).
	
	\bibitem{GNO}
	J.~J.~C.~Nimmo, C.~R.~Gilson and Y.~Ohta,
	Theor.\ Math.\ Phys.\  {\bf 122}, 239 (2000)
	[Teor.\ Mat.\ Fiz.\  {\bf 122}, 284 (2000)].
	
	\bibitem{OoVa}
	H.~Ooguri and C.~Vafa,
	Nucl.\ Phys.\ B {\bf 361} (1991) 469;
	Nucl.\ Phys.\ B {\bf 367} (1991) 83.
	
	\bibitem{Polyakov}
	A.~M.~Polyakov,
	{\it Gauge Fields and Strings},
	(harwood academic, 1987).
	
	\bibitem{MOS}
	N.~Sasa, Y.~Ohta and J.~Matsukidaira,
	J.\ Phys.\ Soc.\ Jap.\  {\bf 67}, 83 (1998).
	
	\bibitem{Shifman}
	M.~A.~Shifman,
	{\it Instantons in Gauge Theories}
	(World Sci., 1994).
	
	\bibitem{Takasaki_CMP}
	K.~Takasaki,
	Commun.\ Math.\ Phys.\  {\bf 94}, 35 (1984).
	
	\bibitem{tHooft_ansatz}
	G.~'t Hooft, {\it unpublished}.
	
	\bibitem{tHooft}
	G.~'t Hooft,
	{\it 50 Years of Yang-Mills Theory}
	(World Sci., 2005). 
	
	\bibitem{Ward}
	R.~S.~Ward,
	Phil.\ Trans.\ Roy.\ Soc.\ Lond.\ A {\bf 315} (1985) 451.
	
	\bibitem{Wilczek}
	F.~Wilczek, ``Geometry and interactions of instantons,''
	in 
	{\it Quark Confinement and Field Theory} (Wiley, 1977) 211.
	
	\bibitem{Yang}
	C.~N.~Yang,
	Phys.\ Rev.\ Lett.\  {\bf 38} (1977) 1377.
	
\end{thebibliography}
\end{document}